\documentclass[epj]{svjour}
\usepackage{graphics}
\begin{document}
\title{An example of the interplay of nonextensivity and dynamics in the description of QCD matter\\
}
\author{Jacek Ro\.zynek\thanks{\emph{e-mail: jacek.rozynek@ncbj.gov.pl}}
 \and Grzegorz Wilk\thanks{\emph{e-mail: grzegorz.wilk@ncbj.gov.pl}}
}                     
\institute{National Centre for Nuclear Research,
        Department of Fundamental Research, Ho\.za 69, 00-681
        Warsaw, Poland}
\date{Received: date / Revised version: date}

\abstract{Using a simple quasi-particle model of QCD matter, presented some time ago in
the literature, in which interactions are modelled by some effective fugacities $z$, we
investigate the interplay between the dynamical content of fugacities $z$ and effects
induced by nonextensivity in situations when this model is used in a nonextensive
environment characterized by some nonextensive parameter $q \neq 1$ (for the usual
extensive case $q=1$). This allows for a better understanding of the role of
nonextensivity in the more complicated descriptions of dense hadronic and QCD matter
recently presented (in which dynamics is defined by a lagrangian, the form of which is
specific to a given model).
} 

\authorrunning{Jacek Ro\.zynek, Grzegorz Wilk}
\titlerunning{An example of interplay of nextensivity and dynamics}
\maketitle

\section{Introduction}
\label{sec:I}

Dense hadronic or QCD matter is typically produced in a nonextensive environment, i.e.,
in situations where the application of the usual Boltzmann-Gibbs statistics is
questionable (cf. \cite{WW,WW1,WW2,WW3,WW4} and references therein for details). Such an
environment can be described by a nonextensive statistics, which is usually taken to be
in the form of Tsallis statistics \cite{T,T1,T2} and is characterized by a parameter of
nonextensivity, $q\neq 1$ (for $q=1$ one recovers the usual Boltzmann-Gibbs statistics).
The sensitivity of models of high density matter to such an environment has been
investigated for some time already (cf. the most recent works on nonextensive versions of
the Walecka \cite{Santos}, Nambu - Jona-Lasinio \cite{JRGW} or other models
\cite{Deppman1,Deppman2,Lavagno}, and references therein). In practice it amounts to
investigating the departure of values of some selected  observables with increasing value
of the parameter $|q-1|$ from their extensive values (obtained for $q=1$). However, since
in the all above mentioned models the interaction is defined by some form of a more or
less complicated lagrangian, this is not a simple task because particles considered
acquire some dynamical masses which implicitly depend (usually in a very complicated
manner) on the nonextensivity parameter $q$ \cite{JRGW}. It would therefore be
interesting and instructive to demonstrate the sensitivity of the calculational scheme
used to the nonextensive environment in a more transparent way.

Such a possibility is provided by a class of phenomenological quasi-particle models (QPM)
in which the interacting particles are replaced by free, noninteracting quasi-particles.
The effects of interaction,  normally defined by some lagrangian (as, for example, in the
Walecka model \cite{W,W1,W2} or in the Nambu - Jona-Lasinio (NJL) model
\cite{NJL,NJL1,NJL2,NJL3}), are in this class of QMP models modelled phenomenologically
by means of some special, temperature dependent, factors called effective fugacities
$z^{(i)}$ \cite{CR1,CR1a,CR2,CRa3,CR3}), the form of which is obtained from fits to the
lattice QCD results (here provided by \cite{LQCD}). In effect the masses of the
quasi-particles are not directly modified by the interaction\footnote{For a comparison of
this approach with other formulations of the QPM see \cite{CR1,CR1a,CR2,CRa3,CR3} and
references therein.}. The corresponding equilibrium distribution function is assumed to
be equal to
\begin{equation}
f^{( i)}_{eq}(x) = \frac{z^{(i)}e\left(-x_i\right)}{1-\xi\cdot z^{(i)}e\left(-x_i\right)} = \frac{1}{\frac{1}{z^{(i)}}e\left( x_i \right) - \xi}, \label{fex}
\end{equation}
where $e(x) = \exp(x)$, $x_i = \beta E_i$ and $\xi = +1$ for bosons and $-1$ for
fermions. One deals here with particles only: massless $u$ and $d$ quarks ($i=q$) for
which $E_q = p$, strange quarks with mass $m$ ($i=s$) for which $E_s = \sqrt{p^2 + m^2}$
and massless gluons ($i=g$) with $E_g = p$. For $z^{(i)}=1$ one deals with a
noninteracting gas of bosons (fermions).

One can also rewrite Eq. (\ref{fex}) in a form identical to that usually used,
\begin{equation}
f^{(i)}_{eq}(\tilde{x}) = \frac{1}{e\left( \tilde{x}^{(i)}\right) - \xi}, \label{fexm}
\end{equation}
with
\begin{equation}
\tilde{x}^{(i)} = \beta E_i - \mu^{(i)}(T) \label{xeff}
\end{equation}
and
\begin{equation}
\mu^{(i)}(T) =  \ln z^{(i)}(T) \label{muz}
\end{equation}
 representing a kind of {\it effective chemical potential}, $\mu^{(i)}$, which depends on temperature
 $T$ and replaces the action of the fugacities $z^{(i)}$\footnote{Note that this $\mu$ contains both the interaction and standard  chemical potential used, for example, by us in our nonextensive Nambu - Jona-Lasinio approach \cite{JRGW}. Therefore $z=1$ corresponds to the case when the standard chemical potential is equal to the confining potential and we have free particles.}.

Such notation suggests the possibility of a straightforward generalization of Eq.
(\ref{fexm}) to the nonextensive case. To this end, following
\cite{JRGW,Santos,Deppman1,Deppman2,JR}, one simply replaces $f^{(i)}_{eq}$ by the
corresponding nonextensive particle occupation numbers:
\begin{equation}
n_q\left( \tilde{x}^{(i)} \right) = \frac{1}{ e_q\left( \tilde{x}^{(i)}\right) - \xi}, \label{enq}
\end{equation}
where the $q$-exponential function is defined as
\begin{equation}
e_q(x) = [ 1 + (q-1)x ]^{\frac{1}{q-1}}\quad{\rm for}\quad x> 0.  \label{eqe}
\end{equation}
Its inverse function is
\begin{equation}
e_{2-q}(-x) = [1 + (1-q)(-x)]^{\frac{1}{1-q}}  \label{duality}
\end{equation}
(known as the {\it dual} $(2-q)$-exponent), i.e.,
\begin{equation}
e_{2-q}(-x)\cdot e_q(x) = 1. \label{qdualq}
\end{equation}
For $q \rightarrow 1$ one returns to the extensive situation with $e_q(x) \rightarrow e(x)$, $e_{2-q}(-x) \rightarrow e(-x)$ and with relation (\ref{qdualq}) replaced by the usual extensive relation, $e(x)\cdot e(-x) = 1$.

In fact, this prescription works without additional restrictions only as long as $x$ (or
$(-x)$) remains positive. This is always true if $\mu^{(i)}(T) \le 0$ (or $z^{(i)} \le
1$, which is the case for the usual extensive situations \cite{CR1}). However, for the
nonextensive $\mu_q^{(i)}(T)$ this is not always true, therefore the above formulas  have
to be supplemented by some additional conditions (discussed in \cite{JR,JRGW}). These
will be presented in more detail together with the results of our investigations in
Sections \ref{sec:qQPM} and \ref{sec:res}.

Note that with $e_q(x)$ defined by Eq. (\ref{eqe}) one has to use the following form of the respective $q$-logarithm functions:
\begin{equation}
\ln_q X = \frac{X^{q-1} - 1}{q - 1} \stackrel{q \rightarrow 1}{\Longrightarrow} \ln X, \label{lnq}
\end{equation}
for which
\begin{equation}
\ln_q \left[ e_q(X) \right] = X. \label{qlne}
\end{equation}
Respectively, with $e_{2-q}(x)$ defined by Eq. (\ref{duality}) one has to use its dual version,
\begin{equation}
\ln_{2-q}X = \frac{X^{1-q}-1}{1-q} \stackrel{q \rightarrow 1}{\Longrightarrow} \ln X, \label{lnq2}
\end{equation}
for which
\begin{equation}
\ln_{2-q} \left[ e_{2-q}(X) \right] = X. \label{qlnedual}
\end{equation}

Note also that because of the above duality properties, the nonextensive version of Eq. (\ref{fex}) (with $z^{(i)}=1$) takes the following form:
\begin{equation}
n_q(x) = \frac{1}{e_q(x) - \xi} = \frac{e_{2-q}(-x)}{1 - \xi e_{2-q}(-x)}. \label{n+n}
\end{equation}
A further consequence of this duality is that the known extensive relation,
\begin{equation}
 n(x) + n(-x) = \xi, \label{nplusnminus}
\end{equation}
 now takes the following {\it dual} form \cite{JR,Biro}:
\begin{equation}
n_q(x) + n_{2-q}(-x) = \xi. \label{dualcor}
\end{equation}

\section{QPM in a nonextensive environment: $q$-QPM}
\label{sec:qQPM}

There are two possible approaches to proceed from the usual extensive QPM to its nonextensive version, the $q$-QPM.
\\

\noindent $(A)$~~The first, seemingly very straightforward, has already been mentioned.
One simply takes  the extensive version of the QPM in the form of Eq. (\ref{fexm}) and
changes $\exp( \dots )$ to $\exp_q( \dots )$. This corresponds to insertion of the
initial extensive system in the nonextensive environment characterised by a
nonextensivity parameter $q$; for $q \rightarrow 1$ one recovers the usual extensive
case. The nonextensive formula for the particle occupation number is in this case given,
for $ q > 1$, by Eq. (\ref{enq}) with
\begin{eqnarray}
\tilde{x}^{(i)} \rightarrow x_q^{(i)} &=& \beta E_i - \mu_q^{(i)};~~ \mu_q^{(i)} =  \ln\left[ z_q^{(i)}(T)\right]. \label{xqi}
\end{eqnarray}
For $q < 1$ it is given by Eq. (\ref{eqe}) with $ (-x) \rightarrow x_q^{(i)}$ defined
above. Note that the effective chemical potentials $\mu^{(i)}$, or fugacities $z^{(i)}$
(cf. Eq. (\ref{muz})), must become effectively $q$-dependent quantities because some part
of the original dynamics is now described by the replacement $e(\dots) \rightarrow
e_q(\dots)$. This fact has other consequences. Namely, in the case when the resulting
$z_q^{(i)}$ exceeds unity and the corresponding $\mu_q^{(i)}$ becomes negative, Eqs.
(\ref{enq}) or (\ref{eqe}) have to be supplemented by conditions ensuring that the
corresponding $q$-exponents are always nonnegative real valued (see Sections 3.2 and 3.3
of \cite{JRGW} for details). As will be seen below, in our case it will result in
$z_q^{(i)}(\tau)$ limited for $q>1$ to some range of $\tau < \tau_{lim}$, such that
$z_q^{(i)}\left( \tau_{lim}\right) = 1$ and the corresponding $q$-exponent becomes zero
forcing the respective particle occupation number to remain equal to unity from this
point \cite{JR,JRGW}\footnote{Because in QMP and in $q$-QMP we do not have a chemical
potential there is also no corresponding Fermi energy. Therefore, the third method of
introducing nonextensivity discussed in Section 3.4 of \cite{JRGW} is not applicable
here.}. Note that in this approach the energies $E_i$ remain unchanged, the only
dynamical change introduced by
switching to a nonextensive environment is in $z^{(i)} \rightarrow z_q^{(i)}$.\\

\noindent $(B)$~~In the second approach one starts with some system of noninteracting
particles and first immerses it in a nonextensive environment characterized by a
nonextensivity parameter $q\neq 1$; they will then be described by Eq. (\ref{enq}) (with
$\tilde{x}^{(i)} = \beta E_i$). The $q$-QPM is then defined by introducing, as before, a
$q$-fugacity factor, $z_q^{(i)}(T)$, and defining particle occupation numbers
as\footnote{Note the important difference between methods $(A)$ and $(B)$. In method
$(A)$ the original fugacity described {\it the interaction of extensive quasiparticles},
whereas in method $(B)$, the $q$-fugacity describes {\it the interaction of nonextensive
quasiparticles}, i.e., quasiparticles in some nonextensive environment.}
\begin{equation}
n_q\left( x_i \right) = \frac{1}{ \frac{1}{z_q^{(i)}}e_q\left( x_i\right) - \xi},\quad x_i = \beta E_i. \label{zenq}
\end{equation}
In this case one can also introduce a $q$-version of the effective chemical potential,
$\mu_q^{(i)}$, and rewrite, for $q>1$, Eq. (\ref{zenq}) as
\begin{equation}
n_q\left[ x_q^{(i)} \right] = \frac{1}{e_q\left[ x_q^{(i)}\right] - \xi} \label{zenq1}
\end{equation}
where now
\begin{eqnarray}
x_q^{(i)} &=& \beta\cdot E^{(i)}_q - \mu_q^{(i)};\quad \mu_q^{(i)} = \ln_{2-q}\left[z_q^{(i)}\right] \label{equality}\\
{\rm and} && E_q^{(i)} = \left[ z_q^{(i)}\right]^{1-q}\cdot E_i. \label{qE}
\end{eqnarray}
As in case $(A)$, for $q < 1$ it is given by Eq. (\ref{eqe}) with $ (-x) \rightarrow
x_q^{(i)}$ defined above. All remarks concerning supplementary conditions needed in this
case are identical to those brought up when presenting approach $(A)$ above. Note that in
this case not only $z^{(i)} \rightarrow z_q^{(i)}$ but also the form of the effective
chemical potential (its dependence on fugacity) is different and the initial energy now
becomes a $q$-dependent quantity as well.

When going into detail we follow closely the approach developed in \cite{Santos,Deppman1}
and take  for the nonextensive ideal quantum gas the following form of the nonextensive
partition function $\Xi_q$ \footnote{In \cite{Santos} it was derived from first
principles using the so-called $q$-calculus, in \cite{Deppman1} it was just postulated.}:
\begin{equation}
\ln_q \left( \Xi_q\right) = - \int \frac{d^3 p}{(2\pi)^3}\sum_i \xi L_q\left[ x_q^{(i)}\right], \label{qXi}
\end{equation}
where the summation is, as in \cite{CR1}, over the type of partons considered, with $i=q$
and $\xi = -1$ for light quarks (for which we assume zero mass), $i=g$ and $\xi = +1$ for
(massless) gluons and $i=s$ and $\xi = -1$ for strange quarks with mass $m$. Functions
$L_q(x)$ are defined as
\begin{equation}
L_q(x) = \ln_{2-q} \left[ 1 - \xi e_{2-q}(-x) \right]. \label{Lq}
\end{equation}
As in \cite{CR1,CR1a,CR2,CRa3,CR3} we do not consider antiparticles.

Eq. (\ref{qXi}) can also be written in a different form (used, for example, in
\cite{Santos}). Integrating by parts one gets
\begin{eqnarray}
&&\int_0^{\infty}\! p^2 dp\, \ln_{2-q}\left[ 1 - \xi e_{2-q}(-x) \right] = \nonumber\\
&&=  -\frac{1}{3} \int_0^{\infty}\! p^3dp \frac{\partial}{\partial p} \left\{ \ln_{2-q}\left[ 1 - \xi e_{2-q}(-x) \right]\right\}. \label{Pnq}
\end{eqnarray}
Because
\begin{eqnarray}
\frac{\partial \ln_{2-q}(x)}{\partial x} &=& \xi \left[ 1 - \xi e_{2-q}(-x)\right]^{-q}\cdot \left[e_{2-q}(-x)\right]^q =\nonumber\\
&=& \frac{\xi}{\left[ e_q(x) - \xi\right]^q} = \xi \left[ n_q(x)\right]^q \label{nq}
\end{eqnarray}
one can write Eq. (\ref{qXi}) as
\begin{equation}
\ln_q \left( \Xi_q\right) = \frac{1}{3} \int \frac{d^3 p}{(2\pi)^3}\sum_i p \left[ n_q\left( x_q^{(i)}\right)\right]^q \frac{\partial x_q^{(i)}}{\partial p}. \label{qXia}
\end{equation}
The form of the variable $x_q^{(i)}$ depends on the particular implementation of $q$-QPM.
In method $(A)$ it is given by Eq. (\ref{xqi}), in method $(B)$ by Eqs. (\ref{equality})
and (\ref{qE}). This means that
\begin{equation}
\frac{\partial x_q^{(g,q)}}{\partial p} = \beta ~ {\rm and}~  \frac{\partial x_q^{(s)}}{\partial p} = \beta \frac{p}{\sqrt{p^2 + m^2}} \label{case1}
\end{equation}
in the first case and
\begin{eqnarray}
\frac{\partial x_q^{(g,q)}}{\partial p} &=& \beta \left[z_q^{(g,q)}\right]^{1-q}~ {\rm and}\nonumber\\
\frac{\partial x_q^{(s)}}{\partial p} &=& \beta \frac{p}{\sqrt{p^2 + m^2}}
\left[z_q^{(g,q)}\right]^{1-q} \label{case2}
\end{eqnarray}
in the second case. The conditions to be satisfied in order to proceed from Eq.
(\ref{qXi}) to Eq. (\ref{qXia}) are the same as those which must be satisfied by $(x,q)$
in Eqs. (\ref{eqe}) and (\ref{duality}) and which were discussed in detail in \cite{JR}.
Note that the correct particle number density when considering the nonextensive case is
given not by $n_q(x)$ but by $n_q^q(x)$. This is also a necessary condition to satisfy
the thermodynamic consistency of our approach, cf. \cite{JRGW}.

\section{Results}
\label{sec:res}

To check the sensitivity of the quasi-particle approach to the nonextensive environment
characterized by nonextensivity parameter $q$ we use, as our input, results for the
scaled temperature dependence of the fugacities, $z^{(i)} = z^{(i)}(\tau)$ (where $\tau
=T/T_c$ and $T_c$ is the critical temperature), obtained in \cite{CR1} in the usual
extensive environment from their fits to the lattice QCD results presented in
\cite{LQCD}.  Because in the version of $q$-thermodynamics used here all thermodynamic
relations are preserved, we can compare the pressures in extensive and nonextensive
environments using, after \cite{CR1}, the usual thermodynamic relation,
\begin{equation}
P_q\beta V = \ln_q\left(\Xi_q\right) \label{P}
\end{equation}
calculated, respectively, for $q=1$ and for $q \neq 1$ cases:
\begin{equation}
P_{q=1}\left( \tilde{x}_i \right) = P_q\left[ x_q^{(i)}\right]. \label{P_Pq}
\end{equation}
Whereas $\tilde{x}_i$ is given by Eq. (\ref{xeff}), the meaning of $x_q^{(i)}\left( x_i\right)$ depends on the version of $q$-QPM used. In version $(A)$ it is given by Eq. (\ref{xqi}), in version $(B)$ by Eq. (\ref{equality}).

\begin{figure*}[h]
\resizebox{0.5\textwidth}{!}{%
  \includegraphics{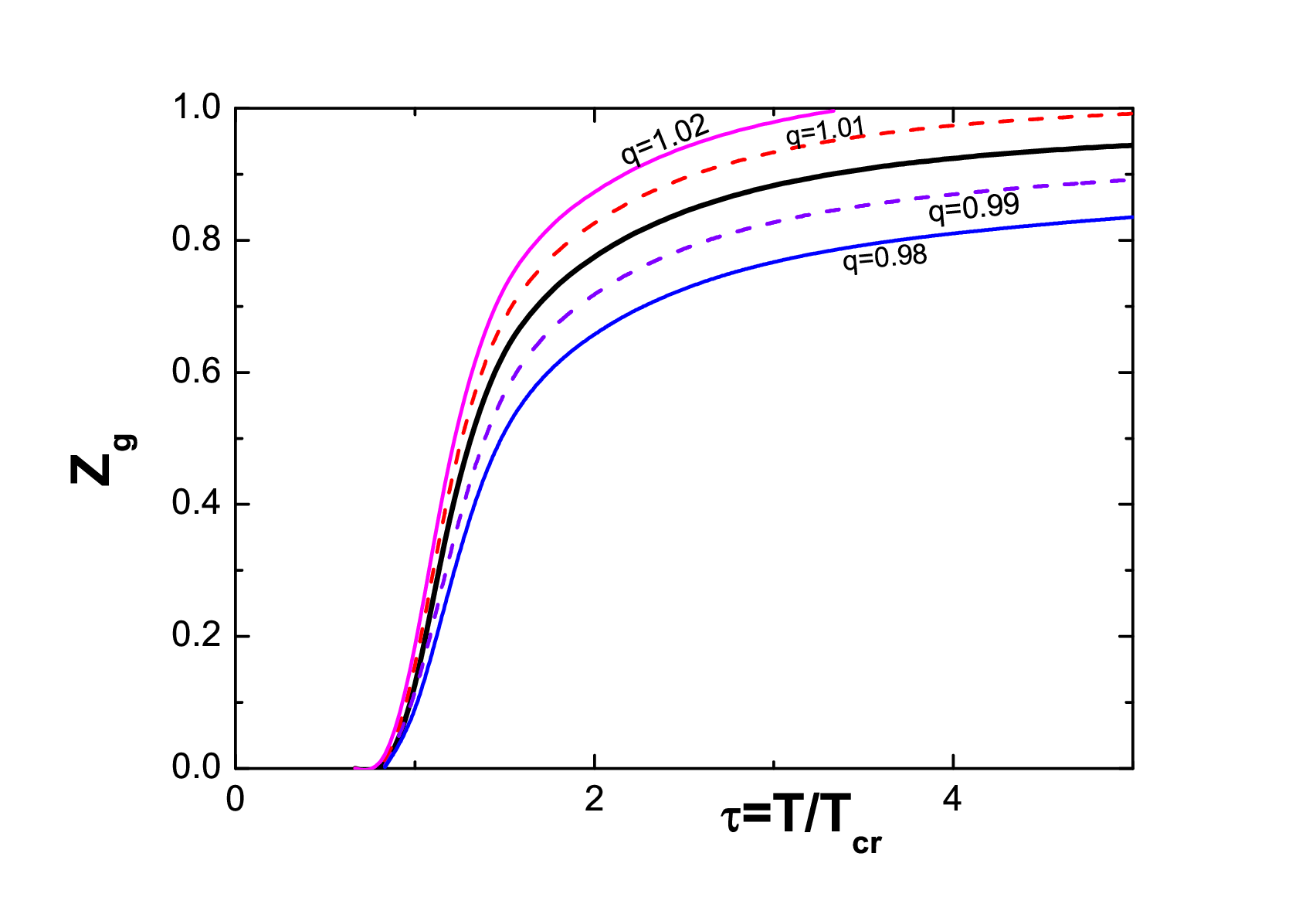}
}\hspace{5mm}
\resizebox{0.5\textwidth}{!}{%
  \includegraphics{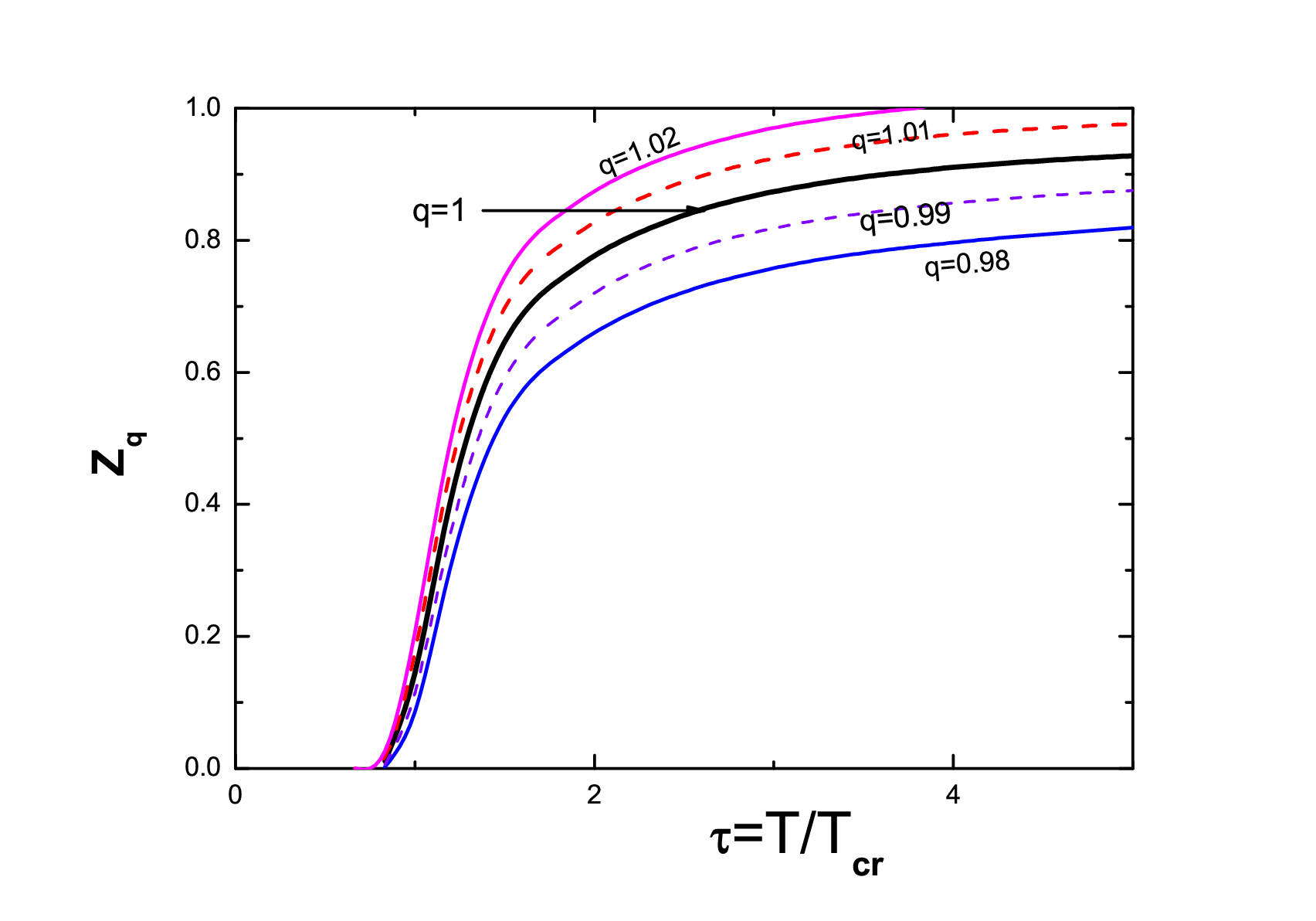}
} \vspace{-4mm}\caption{(Color online) Nonextensive effective fugacities $z_q^{(i)}$ for gluons ($i=g$, left panel) and quarks ($i=q$, right panel) plotted as functions of scaled temperature $\tau = T/T_c$ and obtained in approach $(B)$ (Eqs. (\ref{equality}) and (\ref{qE})). }
\label{z_q}
\end{figure*}
\begin{figure*}[h]
\resizebox{0.5\textwidth}{!}{%
  \includegraphics{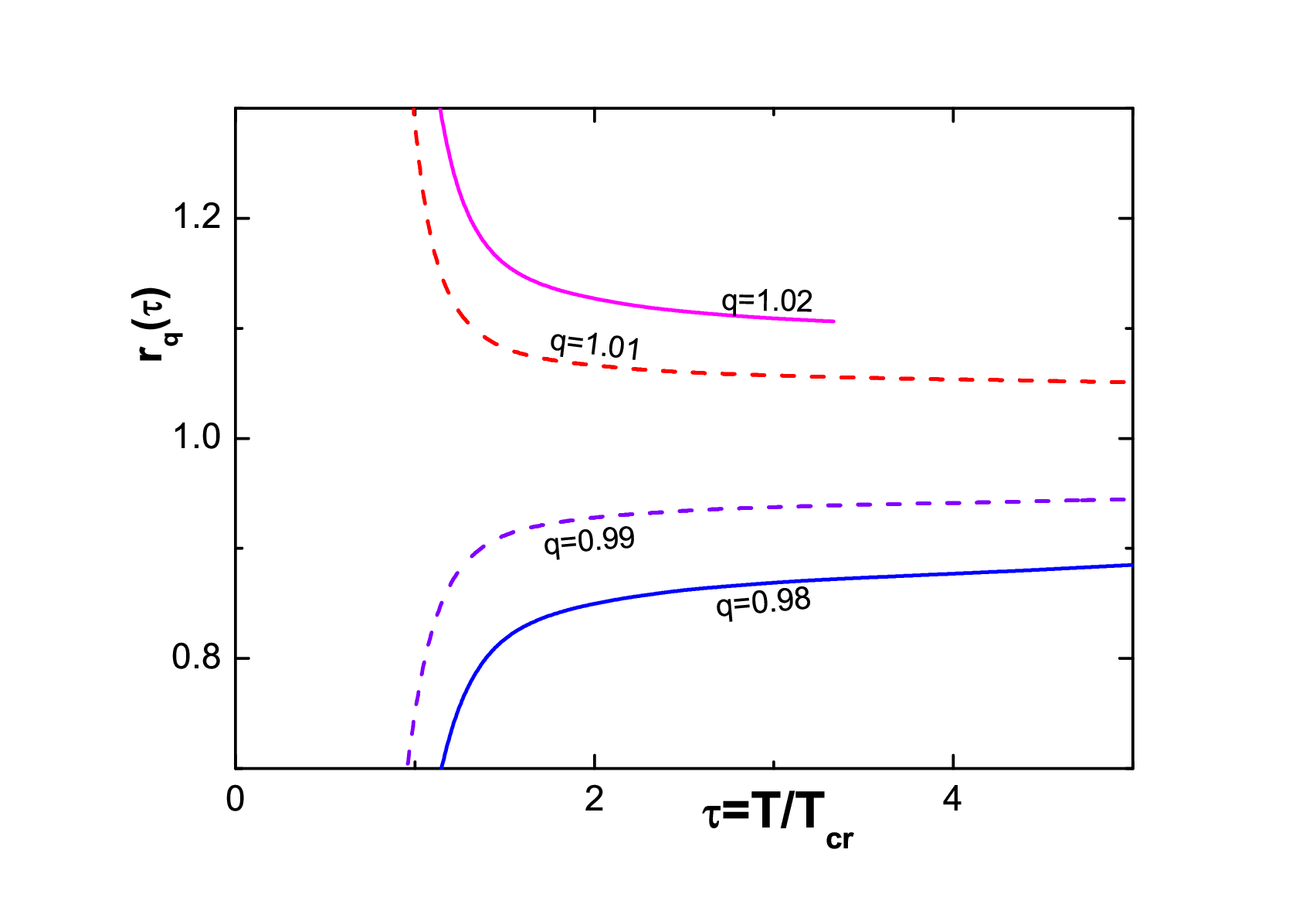}
}\hspace{5mm}
\resizebox{0.5\textwidth}{!}{%
  \includegraphics{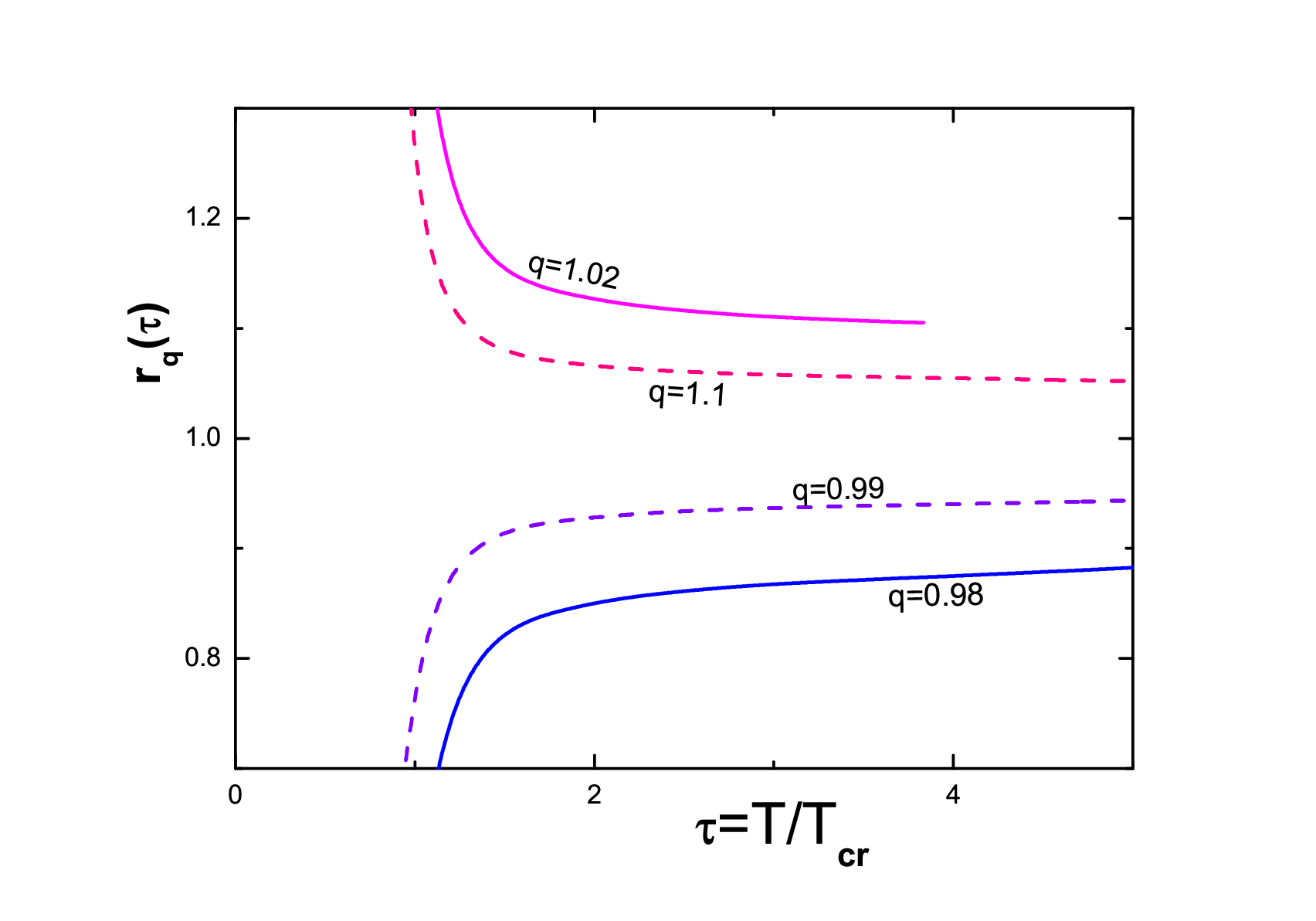}
} \vspace{-4mm}\caption{(Color online)  The same $z_q^{(i)}(\tau)$ as presented in Fig. \ref{z_q} but scaled by their corresponding extensive values (cf. Eq. (\ref{r})). The curves for $q=1.02$ end at $\tau$ for which corrresponding $z_q$ in Fig. \ref{z_q} become unity.}
\label{z_q-r}
\end{figure*}

We are therefore looking for values of the corresponding effective fugacities,
$z_q^{(i)}(\tau)$, which in the nonextensive environment (i.e., on the rhs of Eq.
(\ref{P_Pq})) should replace $z^{(i)}(\tau)$ in the extensive environment(i.e., on the
lhs of Eq. (\ref{P_Pq})) in order to reproduce the  lattice QCD data \cite{LQCD}.
Following \cite{CR1} this is done separately for the gluonic and quark sectors for which
the following conditions must be satisfied:
\begin{eqnarray}
&& \int_0^{\infty} dp p^2 \ln [ 1 - e\left( -\tilde{x}^{(g)} \right)] = \nonumber\\
&&\,\,\,\,\,\,\,\,\,\, = \int_0^{\infty} dp p^2 \ln_{2-q}\left[ 1 - e_{2-q}\left( -x_q^{(g)}\right)\right],\label{gluons}
\end{eqnarray}
for gluons and
\begin{eqnarray}
&& \nu_q \int_0^{\infty} dp p^2 \ln [ 1 + e\left( -\tilde{x}^{(q)} \right)] + \nonumber\\
&& \,\,\,\,\,\,\,\,\,\, +\nu_s \int_0^{\infty} dp p^2 \ln [ 1 + e\left( -\tilde{x}^{(s)} \right)] = \nonumber\\
&& = \nu_q \int_0^{\infty} dp p^2 \ln_{2-q}\left[ 1 + e_{2-q}\left( -x_q^{(q)}\right)\right] +\nonumber\\
&&\,\,\,\,\,\,\,\,\,\, + \nu_s \int_0^{\infty} dp p^2 \ln_{2-q}\left[ 1 + e_{2-q}\left( -x_q^{(s)}\right)\right], \label{quarks}
\end{eqnarray}
for quarks; following \cite{CR1}, $\nu_g = 16$, $\nu_q= 24$ and $\nu_s = 12$. The above
equations provide us with $\tau$ and $q$-dependent relations between the extensive
fugacities, $z^{(i)}(\tau)$ (which are our input), and nonextensive fugacities,
$z_q^{(i)}(\tau)$ (which are our results).

\begin{figure*}[t]
\resizebox{0.5\textwidth}{!}{%
  \includegraphics{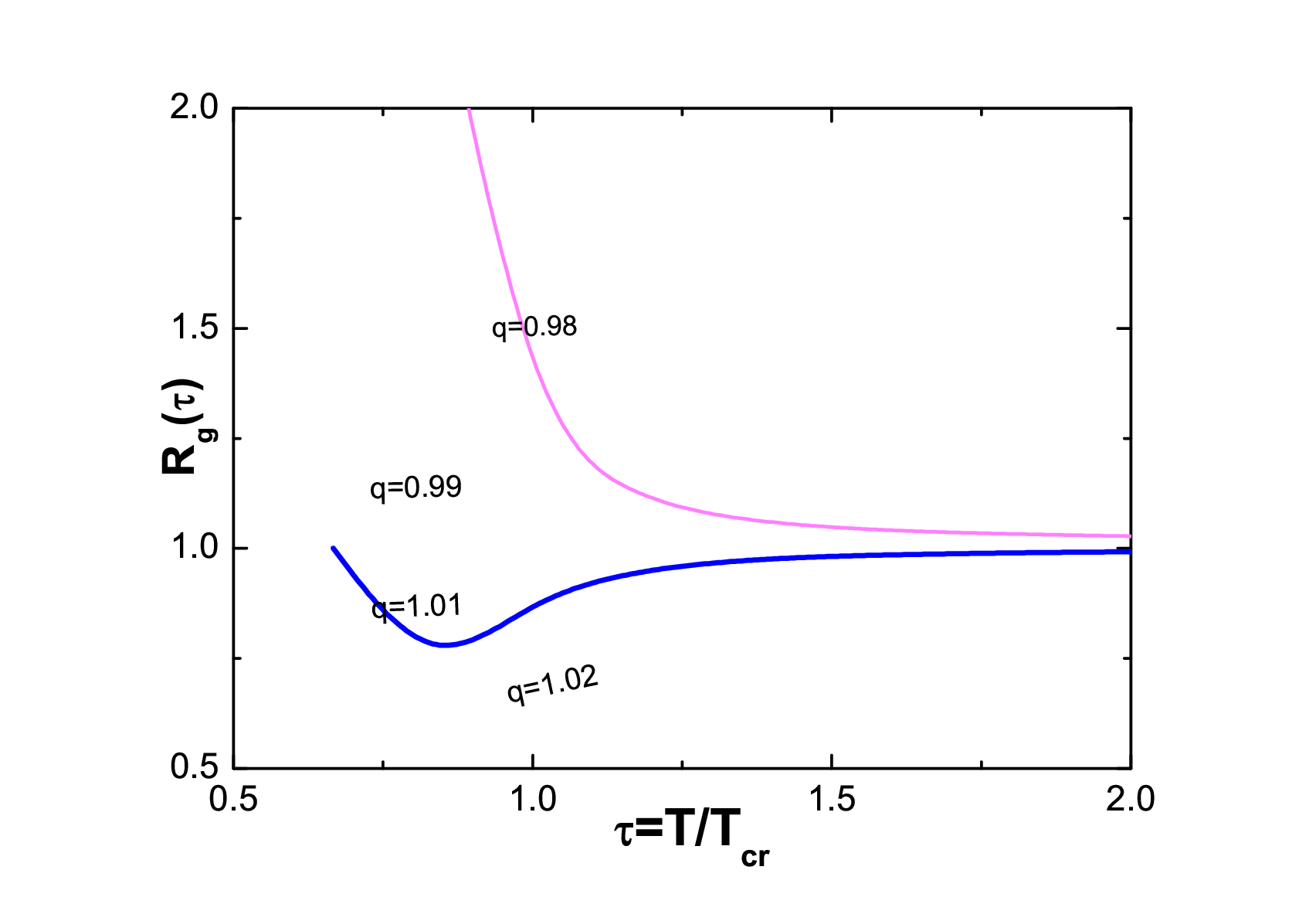}
  }\hspace{5mm}
\resizebox{0.5\textwidth}{!}{%
  \includegraphics{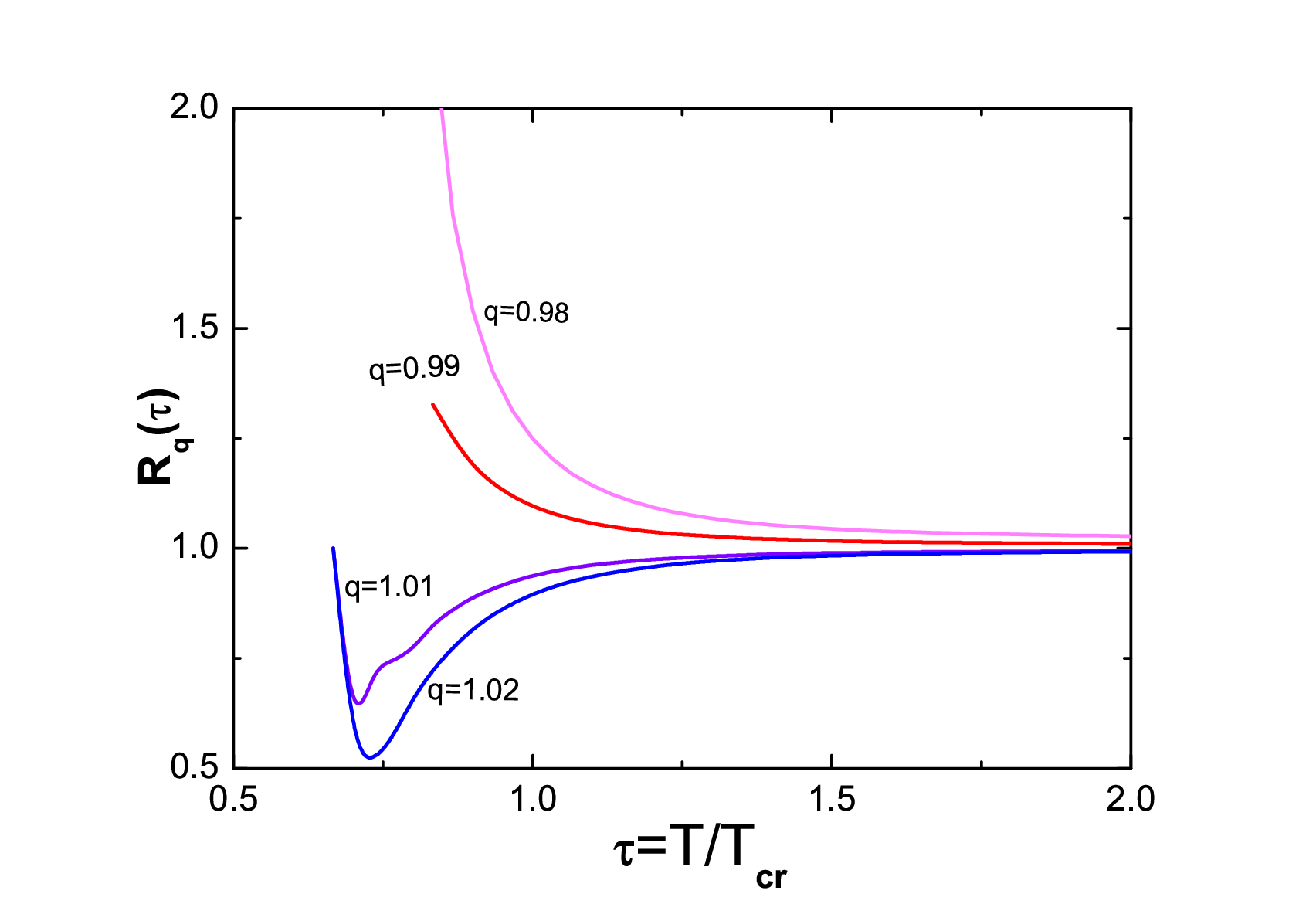}
} \vspace{-4mm}\caption{(Color online) Ratios $R(\tau)$ of gluonic fugacities (left panel) and quarkonic fugacities (right panel) calculated by methods $(A)$ and $(B)$ (cf. Eq. (\ref{comp})) for $q=0.98$, $0.99$, $1.01$ and $1.02$ used above. For greater values of $\tau$ this ratio remains essentially unity.}
\label{comparison}
\end{figure*}

As discussed in detail in \cite{CR1}, there is no one universal function describing the
QCD data in the whole range of scaled temperatures $\tau$ used in fits; the cross-over
point is at $\tau_g = 1.68$ for gluons and $\tau_q=1.7$ for quarks. The low and high
$\tau$ domains require different functional forms (the same occurs for quark and gluon
sectors but with different parameters). Following \cite{CR1} we therefore take as our
input:
\begin{eqnarray}
z^{(g,q)} &=&
a_{(g,q)}\exp\left[ -b_{(g,q)}/\tau^5\right]\cdot\Theta\left( \tau_{(g,q)} - \tau\right) +\nonumber\\
&+& a'_{(g,q)}\exp\left[ -b'_{(g,q)}/\tau^2\right]\cdot\Theta\left( \tau - \tau_{(g,q)}\right)\label{zzz}
\end{eqnarray}
with $\left[ a_{(g)}, b_{(g)}\right] = (0.803, 1.84)$, $\left[ a'_{(g)}, b'_{(g)}\right] = (0.98, 0.94)$ for gluons and $\left[ a_{(q)}, b_{(q)}\right] = (0.81, 1.72)$, $\left[ a'_{(q)}, b'_{(q)}\right] = (0.96, 0.85)$ for quarks.

Fig. \ref{z_q} shows the resulting $z_q{(i)}(\tau)$ (separately for gluons, $i=g$, and
quarks, $i=q$) as functions of scaled temperature, $\tau = T/T_{c}$, calculated for
approach $(B)$.  Fig. (\ref{z_q-r}) shows the same $z_q^{(i)}(\tau)$ but scaled by their
corresponding extensive values, i.e., the ratios
\begin{equation}
 r_i = r_i(\tau) = \frac{z_q^{(i)}(\tau)}{z_{q=1}^{(i)}(\tau)}. \label{r}
\end{equation}
The values of the nonextensivity parameter $q$ used here correspond to values of $q$ used
by us before in the $q$ version of the Nambu Jona-Lasinio model \cite{JRGW}.

The same can be calculated using method $(A)$. However, instead of repeating all the
previous figures we simply present in Fig. \ref{comparison} the corresponding ratios of
results calculated using methods $(A)$ and $(B)$ (separately for gluons and quarks and as
function of scaled temperature $\tau$),
\begin{equation}
R_i = R_i(\tau) = \frac{[z_q^{(i)}(\tau)]_{method(A)} }{ [z_q^{(i)}(\tau)]_{method(B)}}. \label{comp}
\end{equation}
Note the noticeable differences between both methods for smaller values of $\tau$ which
tend to vanish for $\tau \ge 1.5$~\footnote{However, because for small values of the
fugacities both methods start to be numerically unstable, the structures observed below
$\tau \sim 0.75$ are not very reliable.}.

The fugacities $z_q^{(i)}$ obtained above constitute our result. They demonstrate in a
very clear way the action of immersing the QPM in a nonextensive environment with $q\neq
1$. They could be used for any further analysis based on the QPM, for example repeating
the whole analysis of \cite{CR1,CR1a,CR2,CRa3,CR3} for the $q\neq 1$ case. However, this
is not our goal. We shall therefore end this section by presenting the physical
significance of the effective nonextensive fugacities by presenting the corresponding
nonextensive dispersion relations (i.e., single particle energies),
\begin{equation}
\varepsilon_q = - \frac{\partial}{\partial \beta}\left(\Xi_q\right). \label{varepsilon}
\end{equation}
In our case, for the first choice of $q$-QPM (Eqs. (\ref{enq}) and (\ref{xqi})), one gets
\begin{equation}
\varepsilon_q^{(i)} = E_i + T^2\frac{\partial \mu_q^{(i)}}{\partial T} = E_i + T^2\left[ \frac{1}{z_q^{(i)}} \frac{\partial z_q^{(i)}}{\partial T}\right]. \label{DR}
\end{equation}
This means that for this form of the $q$-QPM  extensivity affects only the interaction
term. The quasiparticle energies get some additional contributions from their collective
excitations. Note that this additional term occurs because of the temperature dependence
of the effective fugacities and that it can be interpreted as representing the action of
the gap equation in \cite{JRGW} (but with constant energy $E_i$).

For the second choice of $q$-QPM (Eqs. (\ref{zenq1}) - (\ref{qE})) one gets
\begin{eqnarray}
\varepsilon_q^{(i)} &=& E_q^{(i)} + T^2 \cdot \frac{\partial \ln_{2-q}\left(z_q^{(i)}\right)}
{\partial T} = \nonumber\\
&=& E_q^{(i)} + T^2 \cdot \frac{1}{\left[ z_q^{(i)}\right]^{q}} \frac{\partial z_q^{(i)}}{\partial T}, \label{qdisprel}
\end{eqnarray}
with $E_q^{(i)}$ given by Eq. (\ref{qE}). It means that for this form of the $q$-QPM both the initial energy and the interaction term are modified by effects of nonextensivity. As before, all modifications occur because of the temperature dependence of the effective fugacities and can be interpreted as representing action of gap equation in \cite{JRGW}. However, now this representation is more exact because the energy $E_i$ is modified, cf., Eq. (\ref{qE}), and becomes the $q$-dependent quantity \cite{JRGW}.

\section{Summary and conclusions}
\label{sec:SumCon}

This work illustrates how nonextensive environment (modelled by using $q$-exponentials and methods of nonextensive thermodynamics) changes usual extensive calculations. The quasiparticle model \cite{CR1,CR1a,CR2,CRa3,CR3} used here as basis of our comparison allows for apparently maximal possible separation of effects of the usual dynamics (represented by fugacity $z$) from the effects caused by the nonextensive environment (represented by the nonextensivity parameter $q$). We have limited ourselves to investigation of the respective fugacities in two possible realization of the nonextensive version of the quasiparticle model, the $q$-QPM. They differ by the starting point assumed:
\begin{itemize}
\item in method $(A)$ it is gas of free, noninteracting quasiparticles immersed in extensive environment (i.e., free particles with interaction modelled by some assumed fugacities);
\item in method $(B)$ it is gas of free particles immersed in nonextensive environment\footnote{As a matter of fact, in this case these are really not fully free particles but rather a kind of a noninteracting (because $z_q=1$) $q$-quasiparticles.}.
\end{itemize}
The main results are presented in Fig. \ref{z_q}. It is clearly visible that immersing free quasiparticles in some nonextensive environment described by the nonextensivity parameter $q>1$ considerably accelerates approach to the free (nonextensive) quasipartile limit of $z_q =1$. In the case of $q<1$ nonextensive environment this regime is practically never reached. Considering this result a comment concerning comparison with the similar nonextensive Nambu - Jona-Lasinio results \cite{JRGW} are in order. As shown there, nonextensive effects result, for $q > 1$, in the enhancement of the growth of pressure and entropy observed in the critical region of phase transition from quark matter to hadronic matter in lattice calculations for finite temperature \cite{Bazavov}. As a result, for $q>1$ one reaches earlier the limit of noninteracting particles (albeit still remaining in a nonextensive environment), which corresponds to limit $z_q=1$ here (whereas there is no such transition for the $q <1$ case). Note that such limit is the same for quarks and gluons.

Finally, out of two methods of formulating the $q$-QPM presented here, method $(B)$ seems to be more complete and adequate in what concerns the introduction and description of the nonextensive effects. It can therefore be used further to investigate some more complicated aspects of dense matter in a nonextensive quasiparticle approach.

\vspace{1cm}
\centerline{\bf Acknowledgment}

\vspace*{0.3cm} This research  was supported in part by the National Science Center (NCN) under contract DEC-2013/09/B/ST2/02897. We would like to thank warmly Dr Nicholas Keeley for reading the manuscript.


\begin{thebibliography}{99}

\bibitem{WW} G.~Wilk and Z.~W\l odarczyk, Eur.\ Phys.\ J.\ A {\bf 40}, 299 (2009).

\bibitem{WW1} G.~Wilk and Z.~W\l odarczyk,  Eur.\ Phys.\ J.\ A {\bf 48}, 161 (2012).

\bibitem{WW2} G.~Wilk and Z.~W\l odarczyk, Entropy {\bf 17}, 384 (2015).

\bibitem{WW3} G.~Wilk and Z.~W\l odarczyk, Chaos\ Solitons and Fractals 81, 487 (2015).

\bibitem{WW4} G.~Wilk and Z.~W\l odarczyk, Acta Phys. Pol. B {\bf 46}, 1103 (2015).

\bibitem{T} C.~Tsallis, {\it Introduction to Nonextensive Statistical  Mechanics} (Springer, Berlin,
            2009).

\bibitem{T1} C.~Tsallis, Contemporary Physics, {\bf 55}, 179 (2014).

\bibitem{T2} C.~Tsallis, Acta Phys. Pol. B 46 (2015) 1089.
             For an updated bibliography on this subject, see http://tsallis.cat.cbpf.br/biblio.htm;

\bibitem{Santos}  A.~P.~Santos, F.~I.~M.~Pereira, R.~Silva and J.~S.~Alcaniz, J.\ Phys.\ G {\bf 41},
                  055105 (2014).

\bibitem{JRGW} J.~Ro\.zynek and G.~Wilk, Eur.\ Phys.\ J.\ A {\bf 52}, 13 (2016).

\bibitem{Deppman1} E.~Megias, D.~P.~Menezes and A.~Deppman, Physica\ A {\bf 421}, 15 (2015).

\bibitem{Deppman2} A.~Deppman, J.\ Phys.\ G\ {\bf 41}, 055108 (2014).

\bibitem{Lavagno} A.~Lavagno, D.~Pigato, Physica \ A {\bf 392}, 5164 (2013).

\bibitem{JR} J.~Ro\.zynek,  Physica\ A {\bf 440}, 27 (2015).

\bibitem{W} J.~D.~Walecka, Ann.\ Phys.\ {\bf 83}, 491 (1974).

\bibitem{W1} S.~A.~Chin and J.~D.~Walecka, Phys.\ Lett.\ B {\bf 52}, 1074 (1974).

\bibitem{W2} B.~D.~Serot and J.~D.~Walecka, Adv.\ Nucl. Phys.\ {\bf 16}, 1 (1986).

\bibitem{NJL} Y.~Nambu and G.~Jona-Lasinio, Phys.\ Rev.\ {\bf 122}, 345 (1961).

\bibitem{NJL1} Y.~Nambu and G.~Jona-Lasinio, Phys.\ Rev.\ {\bf 124}, 246 (1961).

\bibitem{NJL2} S.~P.~Klevansky, Rev.\ Mod.\ Phys.\ {\bf 64}, 649 (1992);

\bibitem{NJL3} P.~Rehberg, S.~P.~Klevansky and J.~H\"ufner, Phys.\ Rev.\ C {\bf 53}, 410 (1996).

\bibitem{CR1} V.~Chandra and V.~Ravishankar, Phys.\ Rev.\ D {\bf 84}, 074013 (2011).

\bibitem{CR1a} V.~Chandra, Phys.\ Rev.\ D {\bf 84}, 094025 (2011).

\bibitem{CR2} V.~Chandra and V.~Ravishankar, Phys.\ Rev.\ D {\bf 92}, 094027 (2015).

\bibitem{CRa3} V.~Chandra and V.~Ravishankar1, Eur.\ Phys.\ J.\ C {\bf 59}, 705 (2009).

\bibitem{CR3} V.~Chandra and V.~Ravishankar1, Eur.\ Phys.\ J.\ C {\bf 64}, 63 (2009).

\bibitem{LQCD} M. Cheng et al., Phys. Rev. D {\bf 81}, 054504 (2010).

\bibitem{Biro} T.~S.~Bir\'o, K.~M.~Shen and B.~W.~Zhang, Physica\ A {\bf 428}, 410 (2015).

\bibitem{Bazavov} A.~Bazavov et al., Phys.\ Rev.\ D {\bf 80}, 014504 (2009).

\end{thebibliography}
\end{document}